\documentclass[11pt,a4paper]{article}
 \usepackage[dvips]{graphics}
 \usepackage{graphicx}
 \usepackage{afterpage}
 \usepackage{enumerate}
  \usepackage{longtable}

\newcount\longrefs
\def\aap{\ifnum\longrefs=1 {Astron.\ Astrophys.}\else
                           {A\hbox{\rm \&}A}\fi}
\def\aapr{\ifnum\longrefs=1 {Astron.\ Astrophys.\ Rev.}\else
                            {A\hbox{\rm \&}AR}\fi}
\def\aaps{\ifnum\longrefs=1 {Astron.\ Astrophys.\ Suppl.}\else
                            {A\hbox{\rm \&}A Suppl.}\fi}
\def\aj{\ifnum\longrefs=1 {Astron.\ J.}\else
                          {AJ}\fi}
\def\ao{\ifnum\longrefs=1 {Applied Optics}\else
                           {Appl.\ Opt.}\fi}
\def\aspcs{\ifnum\longrefs=1 {Astron.\ Soc.\ Pacific Conf. Series}\else
                           {ASP Conf.\ Ser.}\fi}
\def\apj{\ifnum\longrefs=1 {Astrophys.\ J.}\else
                           {ApJ}\fi}
\def\apjl{\ifnum\longrefs=1 {Astrophys.\ J. Lett.}\else
                            {ApJ}\fi}
\def\aplett{\ifnum\longrefs=1 {Astrophys.\ J. Lett.}\else
                            {ApJ}\fi}
\def\apjs{\ifnum\longrefs=1 {Astrophys.\ J. Suppl.}\else
                            {ApJS}\fi}
\def\apss{\ifnum\longrefs=1 {Astrophys.\ and Space Science}\else
                            {Astrophys.\ Space Sci.}\fi}
\def\araa{\ifnum\longrefs=1 {Ann.\ Rev.\ Astron.\ Astrophys.}\else
                            {ARA\hbox{\rm \&}A}\fi}
\def\azh{\ifnum\longrefs=1 {Astronomicheskii Zhurnal}\else
                            {Astron.\ Zhur.}\fi}
\def\baas{\ifnum\longrefs=1 {Bull.\ Am.\ Astron.\ Soc.}\else
                            {BAAS}\fi}
\def\bain{\ifnum\longrefs=1 {Bull.\ Astronom.\ Institutes Netherlands}\else
                            {Bull.\ Astr.\ Inst.\ Neth.}\fi}
\def\gca{\ifnum\longrefs=1 {Geochim.\ Cosmochim.\ Acta}\else
                           {Geochim.\ Cosmochim.\ Acta}\fi}
\def\grl{\ifnum\longrefs=1 {Geophys.\ Res.\ Lett.}\else
                           {Geoph.\ Res.\ Lett.}\fi}
\def\iaucirc{\ifnum\longrefs=1 {IAU Circulars}\else
                          {IAU Circ.}\fi}
\def\ip{\ifnum\longrefs=1 {in press}\else
                          {in press}\fi}
\def\jgr{\ifnum\longrefs=1 {J.\ Geophys.\ Res.}\else
                           {J.\ Geophys.\ Res.}\fi}
\def\jrasc{\ifnum\longrefs=1 {J.\ Royal Astron.\ Soc.\ Canada}\else
                           {JRAS Can.}\fi}
\def\mnras{\ifnum\longrefs=1 {Mon.\ Not.\ Roy.\ Astron.\ Soc.}\else
                             {MNRAS}\fi}
\def\nat{\ifnum\longrefs=1 {Nature}\else
                           {Nat}\fi}
\def\pasj{\ifnum\longrefs=1 {Pub.\ Astron.\ Soc.\ Japan}\else
                            {PASJ}\fi}
\def\pasp{\ifnum\longrefs=1 {Pub.\ Astron.\ Soc.\ Pacific}\else
                            {PASP}\fi}
\def\physscr{\ifnum\longrefs=1 {Physica Scripta}\else
                            {Phys.\ Scrip.}\fi}
\def\planss{\ifnum\longrefs=1 {Planetary \& Space Science}\else
                            {Plan. \& Space Sci.}\fi}
\def\procspie{\ifnum\longrefs=1 {Proc.\ SPIE}\else
                            {Proc.\ SPIE}\fi}
\def\qjras{\ifnum\longrefs=1 {Quarterly J.\ Royal Astron.\ Soc.}\else
                            {QJRAS}\fi}
\def\sa{\ifnum\longrefs=1 {Soviet Astron..}\else
                               {Sov.\ Astron.}\fi}
\def\skytel{\ifnum\longrefs=1 {Sky \& Telescope}\else
                            {Sky \& Tel.}\fi}
\def\solphys{\ifnum\longrefs=1 {Solar Phys.}\else
                               {Sol.\ Phys.}\fi}
\def\ssr{\ifnum\longrefs=1 {Space Science Rev.}\else
                               {Space\ Sci.\ Rev.}\fi}

\begin{document}

 \title{\bf Iron abundance in the atmosphere of Arcturus}
\author{\bf V. A. Sheminova}
 \date{}

 \maketitle
 \thanks{}
\begin{center}
{Main Astronomical Observatory, National Academy of Sciences of
Ukraine,
\\Akademika  Zabolotnoho 27,  Kyiv,  03680 Ukraine\\ e-mail: shem@mao.kiev.ua}
\end{center}

 \begin{abstract}

Abundance of iron in the atmosphere of Arcturus has been
determined from the profiles or regions of the profiles of the
weak lines sensitive to iron abundance. The selected lines of Fe I
and Fe II were synthesized with the MARCS theoretical models of
the atmosphere. From the observed profiles of lines available with
a high spectral resolution in the atlas by Hinkle and Wallace
(2005), the values of the iron abundance $A = 6.95 \pm 0.03$ and
the radial-tangential macroturbulent velocity $5.6 \pm 0.2$~km/s
were obtained for Arcturus. The same physical quantities were
found for the Sun as a star; they are $7.42 \pm 0.02$ and $3.4 \pm
0.3$~km/s, respectively. For Arcturus, the iron abundance relative
to the solar one was determined with the differential method as
[Fe/H]~$=-0.48 \pm 0.02$.

\end{abstract}

\section{Introduction}

The abundance of chemical elements in the atmosphere of a star is
traditionally determined from the values of equivalent widths of
spectral lines with the use of completely computer-aided methods.
Such methods allow a large number of lines of different elements
to be quickly processed. They are convenient and widely used in
practice. For example, in the recent paper by Scott et al.
\cite{2015A&A...573A..26S}, new estimates of the abundance of
chemical elements in the solar atmosphere were obtained from the
values of the equivalent widths. However, for many stars, this
approach sometimes yields unreliable results. In those cases when
the stellar spectrum looks like a paling of many lines, a
specified interval of the spectrum is synthesized, and the
profiles of the selected lines rather than the equivalent widths
are analyzed. Such a synthesis requires the data on the atomic
parameters of all of the lines in the interval; however, as a
rule, they are far from always well known. Moreover, not all of
the weak lines have been identified so far. If individual spectral
lines can be isolated in the observed spectrum, the synthesis of
their profiles may yield reliable results in the analysis of the
chemical composition of stars.

As was shown in our previous study \cite{2014MNRAS.443.1967S}, to
determine the abundance, only those regions of the profile that
are most sensitive to it rather than the whole line profile can be
used. Such an approach is especially urgent in those cases when
one wing of the profile is strongly blended, or the intensity in a
red wing of the line is obviously insufficient, or the continuum
cannot be unambiguously determined due to the absorption in many
weak lines, or there is a strong deviation from the local
thermodynamic equilibrium (LTE) in the line core. The new approach
suggested by Sheminova and Cowley \cite{2014MNRAS.443.1967S}
allows the abundance to be accurately obtained when the equivalent
widths of lines and their complete profile cannot be reliably
measured.

The purpose of the present paper is to determine the abundance of
iron in the atmosphere of Arcturus with a high accuracy from those
parts of the profiles of spectral lines that are most sensitive to
the abundance.

\section{Selection of lines for determining the iron abundance}

To determine the abundance as accurate as possible from adjusting
the shapes of the profiles of spectral lines, it is very important
to select proper lines from the stellar spectrum. It is stressed
in  \cite{2014MNRAS.443.1967S} that the weak lines with the
central depths $R=1-{F_{\rm \lambda}}/{F_{\rm c}} \leq 30$\%
(where $F_{\rm \lambda}$ and $F_{\rm c}$ are the radiation fluxes
in the line and continuum, respectively) are most suitable for
determining the abundances. It is well known that the weaker the
absorption line, the less sensitive its profile to the
microturbulent velocity $\xi_{\rm mic}$. As a rule, this parameter
can be estimated with a poor accuracy, and the uncertainty in its
value directly influences the abundance estimate. The tests showed
that the profiles of weak lines with $R$ less than 10\% are
practically indifferent to the change of the velocity $\xi_{\rm
mic}$. They are most suitable for determining the abundance. The
use of the line profiles with the central depths of 10--20\% and
20--30\% may yield the error in the abundance of approximately
$\pm0.01$~dex and $\pm0.02$~dex, respectively, if $\xi_{\rm mic}$
changes by $\pm0.5$~km/s. The profiles of moderate lines ($R$
varies from 30 to 60\%) are most sensitive to $\xi_{\rm mic}$,
because of which their synthesis requires accurate values of the
microturbulent velocity for each of the lines. Note that the
profiles of weak lines are also less sensitive to the damping
effects. Due to this, the selection of lines with $R < 30$\% may
substantially reduce the influence of the ambiguity in the
parameters $\xi_{\rm mic}$ and damping constant.

The profiles of weak lines are substantially influenced by such
parameters as the macroturbulent velocity $\xi_{\rm mac}$, the
rotation velocity of a star $V \sin i$, the abundance of the
element $A$, and the oscillator strengths $\log gf$ of a specified
line. Among the listed parameters, except the abundance, the
parameter $\log gf$ is most crucial. The oscillator strengths
influence the line profiles in the same way as the abundance does,
and it is difficult to separate their effects. The only way to
obtain a correct value of the abundance is to choose the lines for
which this parameter is known as accurately as possible.
Naturally, it is desirable that the oscillators' strengths are
obtained from laboratory experiments.

It is of particular importance that the lines used in the
synthesis have at least one wing or a large part of a wing that is
free of blending. If there are no obvious blends in the wings,
though the line is strongly asymmetric, this asymmetry may be
caused by the influence of very weak blends. Only experience may
show what lines can be used for determining the exact value of the
abundance and what lines cannot.

Thus, to select the absorption lines of Fe I and Fe II, we used
the following criteria:

(1) the depth $R$ in the center of a line should be less than 0.3;

(2) the profiles should contain the regions that are free of
blends;

(3) the parameters $\log gf$ should be measured in laboratory with
a high accuracy.

To determine the value of the iron abundance in the atmosphere of
Arcturus relative to the solar one as exactly as possible, the
same lines in the spectra of Arcturus and the Sun were selected.
In this case, the accuracy in the oscillator strengths will
produce no effect on the relative abundance. First, the lines in
the spectrum of Arcturus were chosen from the spectral atlas of
Hinkle and Wallace \cite{2005ASPC..336..321H}; then, the same
lines were checked in the spectrum of the Sun as a star. The
spectral atlas \cite{2005ASPC..336..321H} contains the normalized
spectrum of the flux of Arcturus and the Sun as a star that was
measured with a high signal-to-noise ratio and a high spectral
resolution (150000 and 300000 for Arcturus and the Sun,
respectively) in the visible range.

To retrieve the value of the absolute abundance of iron in the
atmosphere of Arcturus as accurately as possible, the data on the
oscillator strengths of the weak lines of Fe I were taken from a
paper of Fuhr and Wiese \cite{2006JPCRD..35.1669F} that contains
only the experimental data divided into five ranges by accuracy
($\pm3$, 10, 25, 50, and $>50$\%). We used only the data with the
accuracy of $\pm 3$ and $\pm 10$\%. The oscillator strengths for
the Fe II lines were taken from a paper of Mele'ndez and Barbuy
\cite{2009A&A...497..611M} that contains the most reliable data
currently available.

Note, in the spectrum of Arcturus, many weak lines are partially
blended or subjected to the ``line haze'' effect more strongly
than those in the solar spectrum. To take this into consideration,
at least partly, in the synthesis of the profile of a specified
line, we calculated all of  the blends. Their list, together with
their atomic numbers, was taken from the atomic spectral line
database by R. Kurucz (CD-ROM 23,
http://www.pmp.uni-hannover.de/cgi-bin/ssi/test/kurucz/sekur.html).
Nevertheless, we failed to satisfactorily reproduce the observed
profiles for all of the selected lines. The use of the list of
spectral lines from the Vienna Atomic Line Database (VALD)
\cite{1999A&AS..138..119K} did not allow the problem of lacking
lines to be solved either. This fact shows that the lists of lines
are probably incomplete or the accuracy in the atomic parameters
of lines is insufficient. Consequently, the number of the weak
lines satisfying the above listed requirements turned out to be
not so large.

Table 1 presents a list of the weak lines selected for determining
the iron abundance in the atmospheres of Arcturus and the Sun.


 \begin{table}
 \centering
\caption{\small  Selected weak lines and the values of the iron
abundance obtained for the atmospheres of Arcturus and the Sun  }
\vspace {0.3 cm}
 \label{T:sun}

 \footnotesize
 \begin{tabular}{ccc|cccc |cccc|c}
 \hline\hline
 $\lambda ,$ & $EP,$   & $\log gf$& $R$& $A$ &  $\xi^{\rm RT}_{\rm mac}$, & $\chi^2$, $10^{-5}$& $R$ & $A$  & $\xi^{\rm RT}_{\rm mac}$, & $\chi^2$, $10^{-5}$&[Fe/H]  \\
 nm       & eV     &          &    &     & km/s                    &         &     &      & km/s                   &\\

 \hline
 && &    \multicolumn{4}{c} {Arcturus} &\multicolumn{4}{|c|}{Sun} &\\
 \hline  \hline
\multicolumn{12}{c} {Fe I}\\
 \hline
541.2786 & 4.35 &  -1.72 &0.19   &6.95&  5.82&  0.2&  0.19 &  7.44  &3.35  & 0.2& -0.49 \\
566.1345 & 4.28 &  -1.76 &0.30   &6.89&  5.86&  5.4&  0.20 &  7.39  &3.13  & 0.4& -0.50  \\
585.5077 & 4.61 &  -1.48 &0.25   &6.91&  5.70&  2.9&  0.20 &  7.41  &3.21  & 0.7& -0.52  \\
672.5357 & 4.10 &  -2.10 &0.23   &6.91&  5.69&  6.1&  0.14 &  7.39  &3.32  & 0.2& -0.48 \\
679.3259 & 4.08 &  -2.33 &0.10   &6.95&  5.51&  0.5&  0.10 &  7.42  &3.32  & 0.1& -0.47 \\
680.4271 & 4.58 &  -1.81 &0.15   &6.97&  5.38&  0.1&  0.12 &  7.46  &3.23  & 0.1& -0.49  \\
683.7008 & 4.59 &  -1.69 &0.18   &6.96&  5.46&  0.9&  0.14 &  7.44  &3.11  & 0.6& -0.48  \\
685.4823 & 4.59 &  -1.93 &0.10   &6.96&  5.17&  1.2&  0.10 &  7.48  &3.16  & 0.7& -0.52  \\
 \multicolumn{3}{c|} {Average} & &6.94&  5.57&  2.2&       &  7.43  &3.22  & 0.4& -0.49 \\
          &       &         &      &$\pm$0.03&$\pm$0.24& $\pm$2.4& & $\pm$0.03&$\pm$0.09& $\pm$0.3 &$\pm$0.02 \\

 \hline
\multicolumn{12}{c}{Fe II} \\
 \hline
 472.0149& 3.20  & -4.48  & 0.06 & 6.99 & 5.26 &  0.1&  0.06&7.42 &3.78  &0.4 &-0.43  \\
 608.4111& 3.20  & -3.79  & 0.15 & 6.98 & 5.62 &  0.1&  0.18&7.43 & 3.70 &0.5 &-0.45 \\
 611.3322& 3.22  & -4.14  & 0.09 & 6.99 & 5.26 &  0.1&  0.09&7.44 & 3.62 &0.1 &-0.45 \\
 636.9462& 2.89  & -4.11  & 0.16 & 6.94 & 5.77 &  0.1&  0.16&7.42 & 3.91 &0.3 &-0.48\\
 638.3722& 5.55  & -2.24  & 0.08 & 6.96 & 5.69 &  0.1&  0.07&7.44 & 3.55 &0.1 &-0.48  \\
 643.2680& 2.89  & -3.57  & 0.25 & 6.96 & 5.67 &  5.6&  0.31&7.42 & 3.66 &0.4 &-0.46 \\
 651.6080& 2.89  & -3.31  & 0.30 & 6.94 & 5.82 &  2.1&  0.39&7.42 & 3.66 &0.9 &-0.48  \\
 751.5832& 3.90  & -3.39  & 0.06 & 6.95 & 5.76 &  0.5&  0.09&7.39 & 3.65 &0.1 &-0.44  \\
 \multicolumn{3}{c|} {Average} & & 6.96 & 5.60 &  1.1&      &7.42 & 3.69 & 0.3 &-0.46  \\
         &       &         &        &$\pm$0.02&$\pm$0.22& $\pm$1.8&   &  $\pm$0.02&$\pm$0.11& $\pm$0.3  &$\pm$0.02\\
 \hline
 \multicolumn{3}{c|} {Average for all lines}    &        & 6.95&  5.59& 1.6  &        &  7.42  &3.46  & 0.3  &-0.48 \\
 &       &         &        &$\pm$0.03& $\pm$0.22& $\pm$2.2&        &  $\pm$0.02&$\pm$0.26& $\pm$0.3 &$\pm$0.02 \\
  \hline

 \end{tabular}
 \end{table}
 \noindent


\section{Synthesis of the spectra of Arcturus and the Sun}

The synthesis of spectral lines, which is fundamental for the
analysis of the chemical composition of stars, was performed with
our programming code called SPANSAT \cite{1988ITF...87P....3G}.
The local thermodynamic equilibrium (LTE) is assumed in all of the
calculations. One-dimensional (1D) models of the atmosphere were
built from the MARCS database  of Gustafsson at al.
\cite{2008A&A...486..951G}. For Arcturus, we assumed the effective
temperature $T_{\rm eff} = 4286 $~K, the acceleration of gravity $
\log g = 1.66 $, the metallicity [M/H]~=~$ -0.33 $, and the
abundance of $\alpha$-elements (O, Ne, Mg, Si, S, Ar, Ca, and Ti)
[$\alpha $/Fe] = +0.4 dex, which corresponds to the data of
Ram{\'{\i}}rez and Allende Prieto \cite{2011ApJ...743..135R}. The
reliability of these data has been recently confirmed in our paper
\cite{2013KPCB...29..176S}. To obtain the atmospheric model with
these fundamental parameters, we used six corresponding models
from the MARCS database and interpolated them.

The atmospheric model for the Sun as a star was also taken from
the MARCS database; its parameters are $ T_{\rm eff} = 5777 $~K, $
\log g = 4.44 $, and [M/H]~$ = 0$. The chemical composition of the
solar atmosphere corresponds to the data of Asplund et al.
\cite{2009ARA&A..47..481A}. The rotation velocity was $ V \sin i =
1.5$~km/s \cite{2006PASP..118.1112G} and 1.85 km/s
\cite{1984ApJ...281..830B} for Arcturus and the Sun, respectively.
The van der Waals constant of dumping was calculated with the
Anstee-Barklem-O'Mara method. The required dumping parameters
$\sigma$ and  $\alpha$ were taken from papers
\cite{2005A&A...435..373B, 2000A&AS..142..467B}.

The 1D atmospheric models of Arcturus and the Sun require the
micro- and macroturbulent velocities to be introduced. We
investigated their influence on the line profile in the stellar
spectrum. The most optimal version is the isotropic model of
microturbulence and the radial-tangential model of
macroturbulence. Moreover, the micro- and macroturbulent
velocities do not change with depth in the atmosphere. The
radial-tangential model of macroturbulence suggested by Gray
\cite{1975ApJ...202..148G} is a sum of two Gaussians: $G =
G(\xi^{\rm R}_{\rm mac})S^{\rm R} + G(\xi_{\rm mac}^{\rm T})(1 -
S^{\rm R})$, where $S^{\rm R}$ of the surface area on the disk of
a star occupied by the radial component. We assumed the simplest
case, where $S^{\rm R}=0.5$  and the radial component $\xi^{\rm
R}_{\rm mac}$ is equal to the tangential one $\xi^{\rm T}_{\rm
mac}$. In this case, the velocity of radial-tangential
macroturbulence is usually designated as $\xi^{\rm RT}_{\rm mac}$.
The yielded accuracy of the agreement between the synthesized and
observed profiles is two times higher than that in the isotropic
model of macroturbulence.

\section{Determining the iron abundance}

To determine the iron abundance, the standard fitting procedure of
the synthesized and observed profiles was used. The profiles or
their abundance-sensitive regions should be free of blending. The
free parameters were only the abundance $A$ and the macroturbulent
velocity $\xi_{\rm mac}^{\rm {RT}}$. The remaining parameters were
fixed. According to Ram{\'{\i}}rez and Allende Prieto
\cite{2011ApJ...743..135R}, the microturbulent velocity for
Arcturus is $\xi_{\rm mic} = 1.74$~km/s, while the standard value
of $\xi_{\rm mic} = 1$~km/s was chosen for the Sun as a star in
accordance with the value assumed in the MARCS model of the solar
atmosphere. The Doppler shift of the observed lines relative to
the standard wavelength was determined for each of the lines. The
observed (obs) profiles were fitted with the synthesized (syn)
ones by minimization of  $\chi^2= \frac{1}{N  } \sum_{k=1}^N
(R_{\rm k, syn}-R_{\rm k,obs})^2$, where $N$  is the number of
profile points chosen for fitting. The procedure was automatically
performed with visual checking. The final result (in our case, the
iron abundance, macroturbulent velocity, and the minimization
accuracy) depends on the initial values chosen for free
parameters. This choice is of particular importance, since it
influences the accuracy of determining the abundance. The closer
the initial values to the supposed actual ones, the higher the
fitting accuracy. Because of this, it is necessary to find the
optimal values of the free parameters $A$, and $\xi_{\rm mac}^{\rm
{RT}}$ for each of the lines. Note that these free parameters
produce different effects on the line profile. Our experience
showed that $ \xi_{\rm mac}^{\rm {RT}} $ mostly influences the
external wings of lines, while $A$ influences the internal ones.
From this, we may suppose that the well-chosen initial conditions
will yield the reliable fitting by two free parameters.

In those cases, when the continuum contains very weak unidentified
blends that cannot be taken into account in the synthesis, the
fitting of the profiles in the wing is restricted to approximately
5\%. One more cause of probable uncertainties in the fitting is
the asymmetry of profiles. Sometimes we had to choose only one
wing (more frequently, a blue one) of the line for fitting. Our
experience showed that the blue wing is less deflected by
photospheric convective motions than the red one. Gehren et al.
\cite{2001A&A...380..645G} also noted that the intensity in a red
wing of the lines is often insufficient (the red deficit of
intensity). This fact can be noticed when comparing the observed,
usually asymmetric, profiles to the symmetric synthesized profiles
obtained in the frames of the 1D atmospheric models. The observed
line profile results from the averaging by space and time. It can
be hypothesized that the lines formed in different parts of the
disk, where the temperatures and velocities along the line of
sight are different, will differ in shift, width, and shape.
Because of this, the resultant profile, which is averaged over the
disk, will be asymmetric and shifted by the wavelength. The shape
and the shift of the profile may also be influenced by other
factors, such as spots, oscillations, and pulsations. However, the
main cause of the asymmetry of absorption lines in the stars of
the FGK spectral classes is the convection penetrating into the
photosphere. The influence of convective motions on the profile
was studied in detail in many papers   \cite{2002ApJ...567..544A,
 1981A&A....96..345D, 2002KFNT...18...18S}, explaining how and why the
asymmetry of lines and the red deficit of intensity appear.

Figures 1 and 2 show the results of adjusting the synthesized and
observed regions of the profiles for some lines of Fe I and Fe II.
As is seen, the differences between the synthesized and observed
profiles within the fitted intervals are less than 1\%. At the
same time, for the whole profile, these differences often exceed
1\% due to either the very weak unidentified lines near the
continuum or the inaccurate atomic parameters of the identified
blends or the asymmetry of the observed profiles caused by
convective motions.
 \begin{figure}[t]
 \centerline{
 \includegraphics   [scale=1]{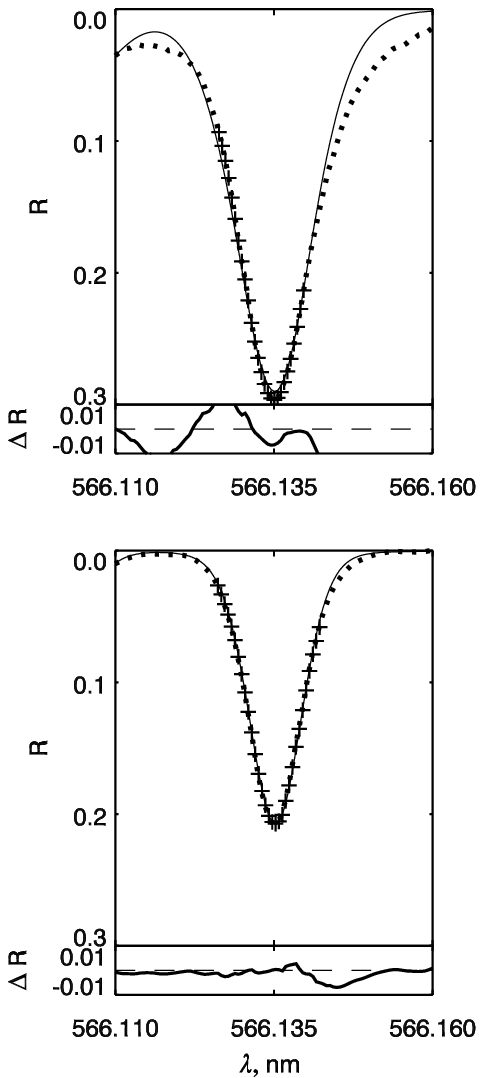}
  \includegraphics  [scale=1]{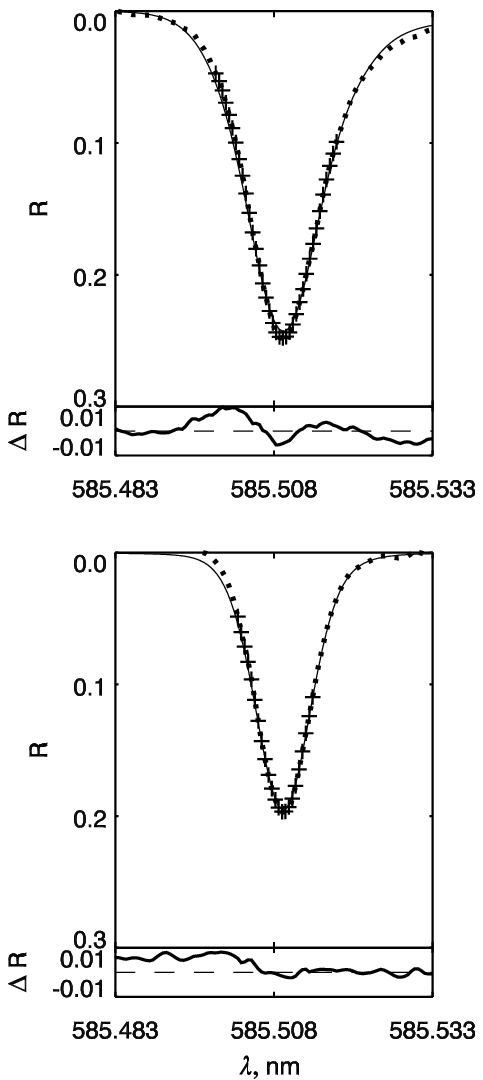}
   \includegraphics [scale=1]{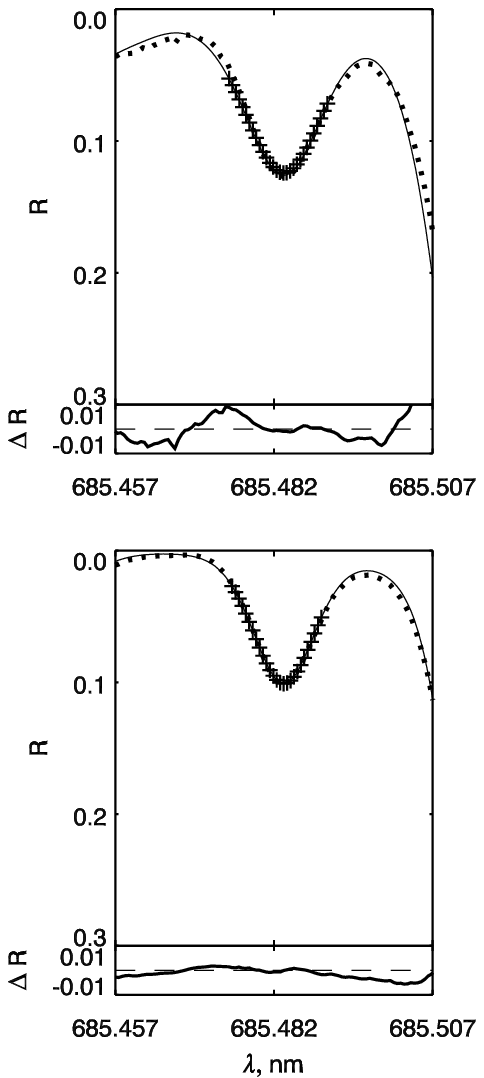}
  }
  \caption {\small
Examples of the best fitting of the profiles $R$ of the Fe I lines
(dotted curves) observed in the spectrum of Arcturus (upper row)
and the Sun (lower row) by the synthesized profiles (solid
curves). The regions chosen for fitting in the observed profiles
are marked by crosses. The differences between the synthesized and
observed profiles
 $ \Delta R=R_{\rm syn}-R_{\rm obs}$
are shown in the lower parts of the plots.
 }
\label{prof1}
 \end{figure}

 \begin{figure}[t]
 \centerline {
 {\includegraphics   [scale=1]{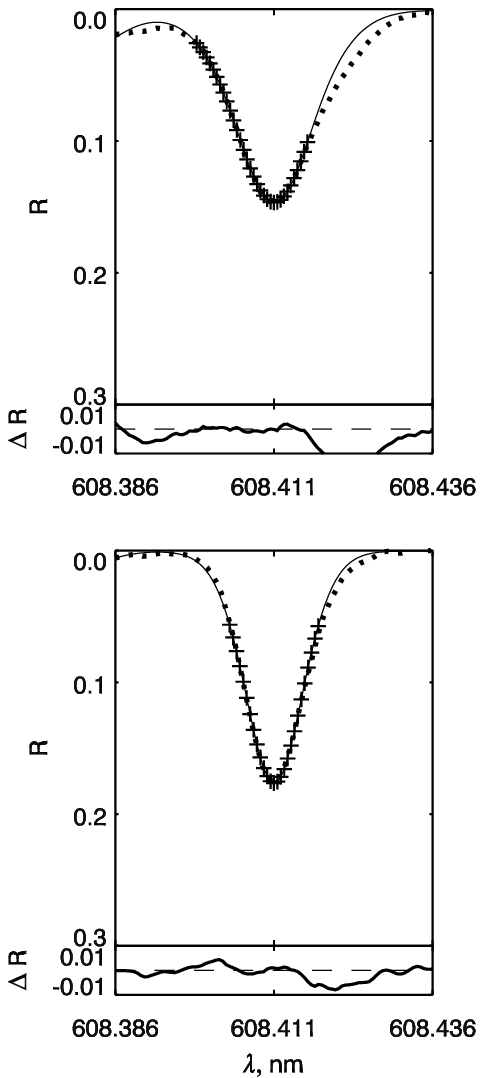}}
 {\includegraphics   [scale=1]{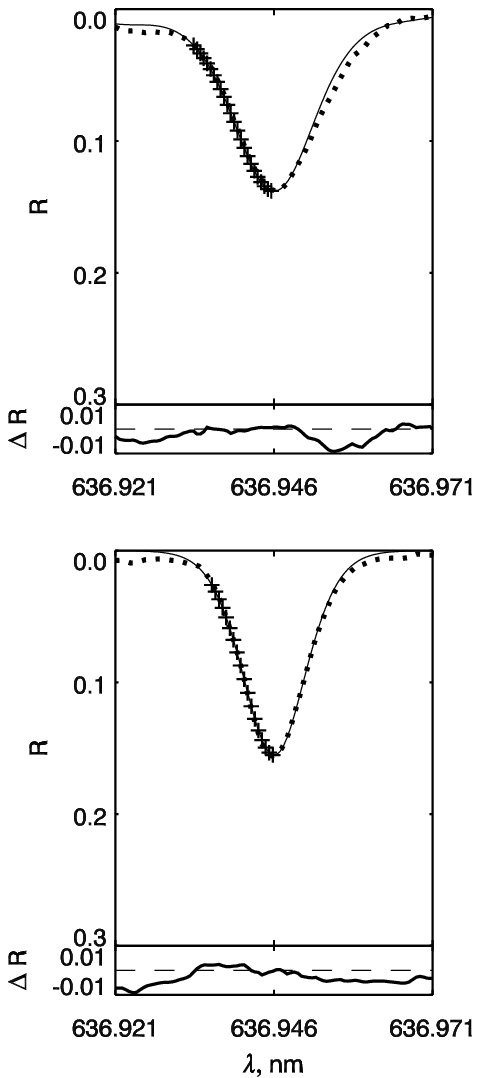}}
  {\includegraphics  [scale=1]{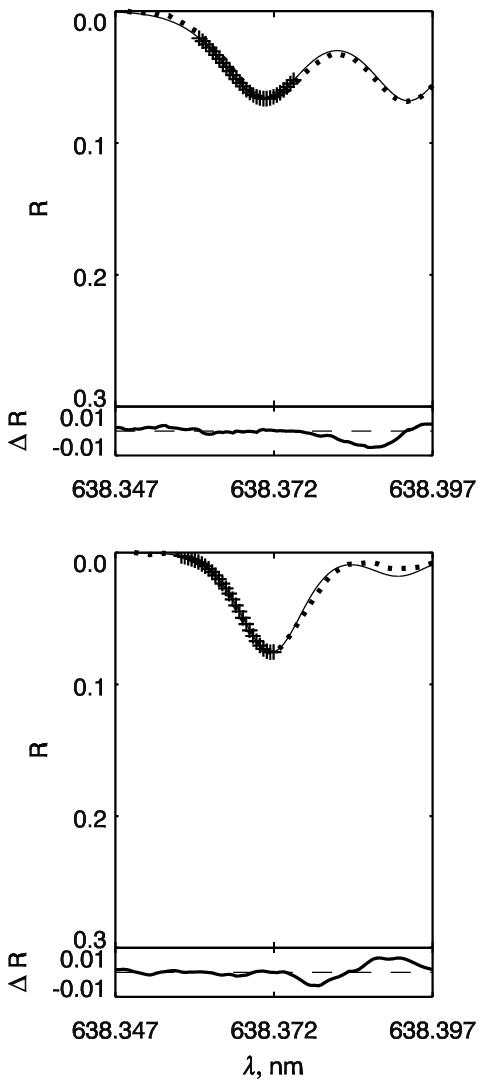}}
    }
  \caption {\small As in Fig. I, except for the Fe II lines.
  } \label{prof2}
 \end{figure}
 \begin{figure}[t]
\centerline{\includegraphics    [scale=0.8]{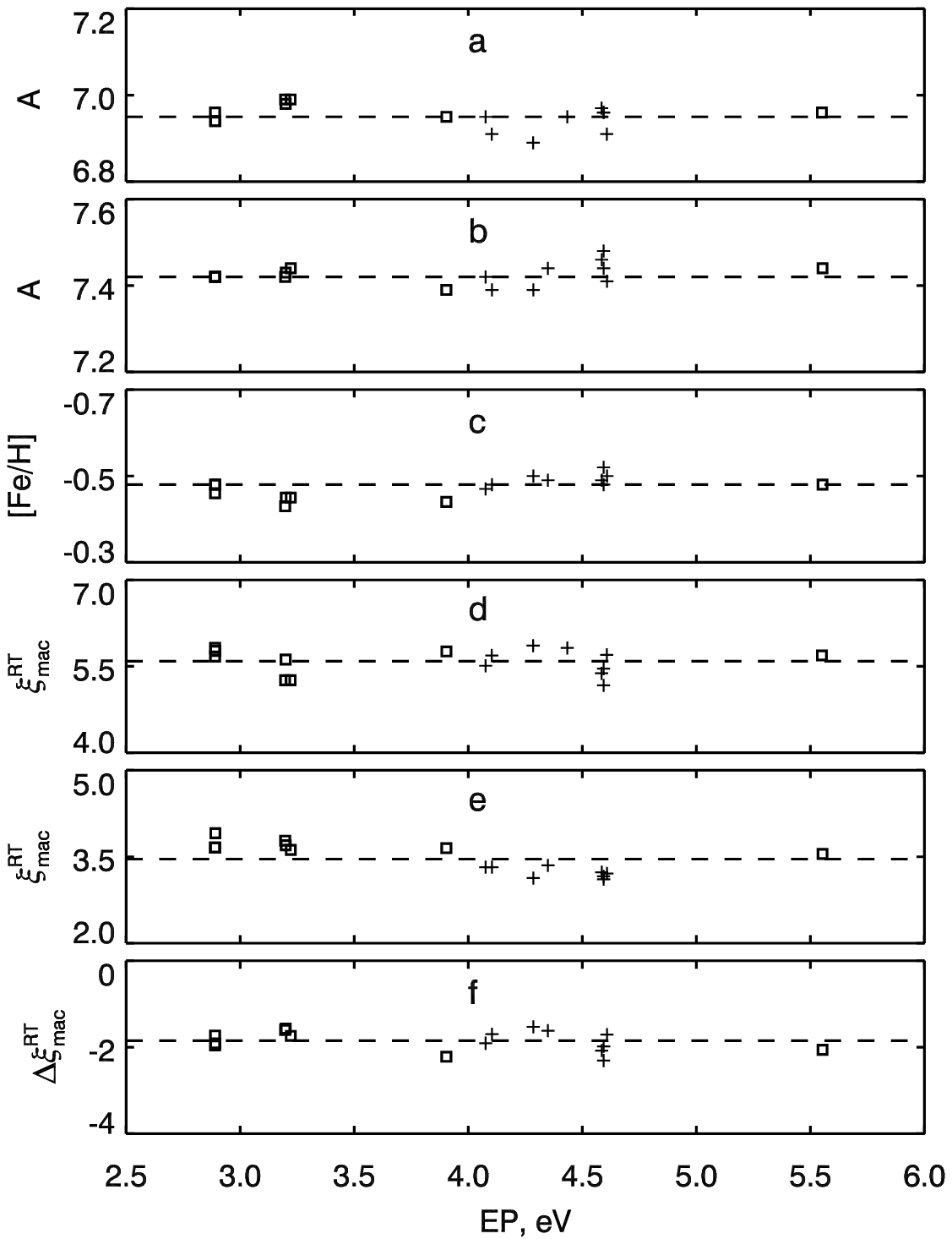}}
 \caption {\small
Individual values of the iron abundance $A$ obtained from the Fe I
(crosses) and Fe II (squares) lines in dependence on the
excitation potential EP for (a) Arcturus and (b) the Sun, (c) the
abundance [Fe/H] for Arcturus relative to the Sun, the
macroturbulent velocities $\xi_{\rm mac}^{\rm {RT}}$ obtained from
the weak lines of Fe I (crosses) and Fe II (squares) for (d)
Arcturus and (e) Sun, and (f) the macroturbulent velocities
$\xi_{\rm mac}^{\rm {RT}}$ for Arcturus relative to the Sun. The
average values for all of the quantities are shown by dashed
lines.
 } \label{abund}
 \end{figure}
The results of determining the iron abundance for each of the
lines in the atmosphere of Arcturus are presented in Table 1. We
obtained the mean value of the iron abundance for Arcturus as
$A_{\rm Fe}=12+\log(N_{\rm Fe}/N_{\rm H}) =6.94 \pm 0.03$ and
$6.96 \pm 0.02$ by the Fe I and Fe II lines, respectively. On
average, the fitting accuracy in the abundance-sensitive regions
of the profiles of the Fe I and Fe II lines is $\chi^2 =2.2 \cdot
10^{-5}$ and $1.1 \cdot 10^{-5}$, respectively. The mean value of
the relative abundance of iron (or metallicity) was obtained as
the difference between the values of the iron abundance in the
atmospheres of Arcturus and the Sun retrieved from the same lines;
it is [Fe/H]~$= A_{\rm Fe}^{\star}-A_{\rm Fe}^{\odot}=-0.49 \pm
0.02$ and $-0.46 \pm 0.02$ by the Fe I and Fe II lines,
respectively.

Figures 3a and 3c present the individual values of the iron
abundance and metallicity for Arcturus obtained by each of the
lines. The average values from all of the Fe I and Fe II lines for
the iron abundance $A=6.95 \pm 0.03$ and [Fe/H]~$=-0.48 \pm 0.02$
are shown by dashed lines.

The values of the radial-tangential macroturbulent velocity in the
atmosphere of Arcturus obtained from adjusting the synthesized and
observed profiles of the selected lines are shown in Fig. 3d. A
dashed line indicates the average value of the macroturbulent
velocity $ \xi_{\rm mac}^{\rm RT}= 5.59 \pm 0.22$~km/s, while the
values of $5.57 \pm 0.25$ and $5.60 \pm 0.22$ were obtained from
the Fe I and Fe II lines, respectively. Figure 3f presents the
values of velocity relative to the solar ones, i.e., $\Delta
\xi_{\rm mac}= \xi_{\rm mac}^{\star} -\xi_{\rm mac}^{\odot}$.

The analogous calculations were performed for the Sun (Table 1).
The mean values of the abundance obtained from the Fe I and Fe II
lines are $A = 7.43 \pm 0.03$ and $7.42 \pm 0.02$, respectively,
and the values of the macroturbulent velocity are $ \xi_{\rm
mac}^{\rm RT}=3.22\pm 0.09$ and $3.69 \pm 0.11$ km/s,
respectively. The fitting accuracy in the abundance-sensitive
regions of the profiles averaged $\chi^2 =0.4 \cdot 10^{-5}$ and
$0.3 \cdot 10^{-5}$ for the Fe I and Fe II lines, respectively.
These values are almost an order of magnitude higher than those
for Arcturus, because the spectral resolution of the solar
spectrum is two times higher. Figure 3b shows the individual
values of the abundance obtained from each of the lines in
dependence on the excitation potential EP. The individual values
of the macroturbulent velocity are shown in Fig. 3e. A dashed line
indicates the mean value of the macroturbulent velocity $ \xi_{\rm
mac}^{\rm RT}=3.46\pm 0.26$~km/s.

\section{Discussion}

In general, the estimates obtained in the present paper
satisfactorily agree with the recently published values of the
iron abundance for Arcturus and the Sun (Tables 2, 3). All the
data for Arcturus (except our estimates) presented in Table 2 were
obtained with the methods based on the adjustment of the
equivalent widths of the Fe I and Fe II lines within the
atmospheric models from the database of Kurucz and their different
modifications. The values of [Fe/H] were found with the
differential method, except those from a paper of Gustafsson et
al. \cite{2008A&A...486..951G}, where the value of $A = 7.45$ was
used for the Sun. As is seen, the parameters of the used models
weakly differ from those of our model. For example, the parameters
of the models of Ram{\'{\i}}rez and Allende Prieto
\cite{2011ApJ...743..135R}  completely coincide with ours, though
their models differ from ours in the same degree as the models by
Kurucz differ from the MARCS models. It is known from a paper
\cite{2008A&A...486..951G} that these models are practically
analogous. The MARCS models are somewhat cooler, and the maximum
at 80 K is for $\tau < 0.01$.

Ram{\'{\i}}rez and Allende Prieto \cite{2011ApJ...743..135R} found
the value of [Fe/H]~$ = -0.52 \pm 0.02 $ and $-0.40 \pm 0.03$ from
the lines of neutral iron and ions, respectively. They recommend
the final value $-0.52 \pm 0.04$. They additionally tested their
result with the MARCS model (i.e., the same as ours) and obtained
$-0.54 \pm 0.05$. This means that the difference in the abundances
obtained with different atmospheric models of Arcturus is 0.02,
while the difference from our result ([Fe/H]~$=-0.48 \pm 0.02$)
within the same model is substantially higher (0.06). We suppose
that, as compared to the method of equivalent widths, the use of
the method of sensitive points favored, first, the decrease of
$\sigma$ (the root mean square deviation) by 0.03 dex and, second,
the increase of the relative value of the iron abundance by 0.06
dex.

In addition, we will pay attention to the known fact (especially
for the Sun) that the results for the iron abundance obtained from
the lines of neutral and ionized iron are different. Rather large
differences can be seen in the data of the other studies (Table
3), from 0.05 to 0.12. In our case, this difference is 0.03. The
large difference can be caused by both the effects connected with
the deviation from LTE and the differences in the models for the
temperature distribution in the photosphere. Note that our method
used weak lines, which allowed us to avoid the errors caused by
the LTE approximation.

The reliability of our results is also evidenced by the fact that
the individual values of [Fe/H] obtained for each of the lines
show no substantial dependence on the excitation potential EP
(Fig.~3c). This means that the chosen atmospheric model is correct
and the lines demonstrate no noticeable deviations from LTE, i.e.,
the lines with high and low excitation potentials yield close
values. We tend to believe that the value we obtained for the iron
abundance in Arcturus relative to the Sun [Fe/H]~$ = -0.48 \pm
0.02$ contains no considerable systematic errors in the frames of
the MARCS model of the atmosphere.

Our purpose was also to obtain, as accurately as possible, the
absolute value of the iron abundance for Arcturus in the usual
hydrogen scale. Such data are not so widely available in the
literature. For the recent years, we have found the data on the
absolute abundance of iron that slightly differ because of the
difference in the list of spectral lines selected for analyzing
the abundance in these studies \cite{2006ApJ...636..821F,
2007ApJ...661.1152F}. As is seen from Table 2, the agreement with
our data is rather good, within the error of analysis. We obtained
the mean value from all of the lines $ A= 6.95 \pm 0.03$; the
values of $\Delta A =A {\rm(Fe~I)} -A{\rm(Fe~II)}=-0.02$ and
$\sigma$ are substantially smaller than those in papers
\cite{2006ApJ...636..821F, 2007ApJ...661.1152F}. This confirms the
reliability of our data.

%
 \begin{table}
  \caption{\small
Values of the iron abundance in the atmospheres of Arcturus
obtained after 2006 from the LTE analysis of the Fe I and Fe II
lines. $T_{\rm eff}/\log gf/\xi_{\rm mic}$ are the parameters used
in the 1D theoretical models of the atmosphere
 }
 \vspace {0.5 cm}
  \label{T:comp2}
  \centering
  \footnotesize
  \begin{tabular}{ccccll}
  \hline
[Fe/H](Fe I)& $A$(Fe I) &[Fe/H](Fe II)& $A$(Fe II) & $T_{\rm eff}/\log
gf/\xi_{\rm mic}$&  Source  \\
  \hline
   $ -0.50\pm0.07$&$6.95\pm 0.07$ &$ -0.55\pm0.05$&$6.99\pm 0.05$ & 4285/1.55/1.61  &  {Fulbright et al. \cite{2006ApJ...636..821F}} \\
  $  -0.54\pm0.04$&$6.91\pm 0.04$ &$ -0.45\pm0.11$&$7.00\pm 0.11$ &4285/1.55/1.62  &  {Fulbright et al. \cite{2007ApJ...661.1152F}} \\
 $ -0.52\pm0.02$&   --   & $ -0.40\pm0.03$&  --    &  4286/1.66/1.74& {Ram{\'{\i}}rez et al. \cite{2011ApJ...743..135R}} \\
  $ -0.49\pm0.07$&  --    & $ -0.40\pm0.04$&  --    &  4290/1.60/1.60 &{McWilliam et al. \cite{2013ApJ...778..149M}} \\
 $ -0.49\pm0.02$&$6.94\pm 0.03$ &$ -0.46\pm0.02$&$6.96\pm 0.02$ & 4286/1.66/1.74   & This study \\
   \hline
 \end{tabular}
 \end{table}
 \noindent

%
 \begin{table}
  \caption{ \small
Iron abundance obtained after 2009 from the analysis of the Fe I
and Fe II lines for the Sun as a star \cite{2009MmSAI..80..643C,
2012MNRAS.427...27B, 2011A&A...528A..87M} and for the center of
the solar disk \cite{2009A&A...497..611M, 2015A&A...573A..26S} }
\vspace {0.5 cm}
  \label{T:comp1}
  \centering
  \footnotesize
  \begin{tabular}{cclll}
  \hline
  \vspace {0.1 cm}
 $A$ (Fe I) & $A$(Fe II) &Atmospheric model & Analysis &  Source  \\
  \hline
  --            & $7.42\pm0.03$  & 1D  MARCS  & LTE& {Melendez \& Barbuy \cite{2009A&A...497..611M}} \\
 $7.45\pm0.06$  & $7.52\pm0.06$  & 3D CO5BOLD &LTE& {Caffau et al. \cite{2009MmSAI..80..643C}} \\
 $7.56\pm0.09$ &   $7.47\pm0.05$&  1D MAFAGS-OS &non-LTE & {Mashonkina et al. \cite{2011A&A...528A..87M}}\\
 $7.43\pm0.05$ &  $7.45\pm0.04$ &   1D  MARCS &LTE& {Bergemann at al. \cite{2012MNRAS.427...27B}}\\
 $7.44\pm0.05$ &  $7.44\pm0.04$ & 1D  MARCS &non-LTE& {Bergemann at al. \cite{2012MNRAS.427...27B}}\\
 $7.48\pm0.02$ & $7.43\pm0.02$  &   1D  MAFAGS-OS &non-LTE& {Bergemann at al. \cite{2012MNRAS.427...27B}}\\
 $7.45\pm0.04$ & $7.51\pm0.04$ &  3D  HD     &LTE& {Scott at al. \cite{2015A&A...573A..26S}}\\
 $7.41\pm0.04$ & $7.42\pm0.03$ &  1D  MARCS  &LTE& {Scott at al. \cite{2015A&A...573A..26S}}\\
 $7.52\pm0.05$ & $7.46\pm0.03$ &  1D  HM     &LTE&{Scott at al. \cite{2015A&A...573A..26S}}\\
 $7.43\pm0.02$ &$7.42\pm0.02$ & 1D  MARCS & LTE&This study \\
   \hline
 \end{tabular}
 \end{table}
 \noindent

With the MARCS model, we also analyzed the iron abundance for the
Sun as a star and obtained $A = 7.43 \pm 0.03$ and $7.42 \pm 0.02$
from the same lines of Fe I and Fe II, respectively. We used the
same method of sensitive points. The value of the abundance
averaged over all lines was $7.42 \pm 0.02$. This analysis was
required in order to derive, as exactly as possible, the relative
abundance for Arcturus [Fe/H] with the differential method, i.e.,
line by line. From the comparison of our results with the others
obtained for the Sun as a star, one may notice that the value of
the iron abundance for the Sun still substantially varies (from
7.41 to 7.56) in spite of a large number of new researches (some
of them are presented in Table 3). In principle, it is clear that
such results are connected with some differences in different
procedures analyzing the iron abundance. The main differences may
arise due to the following causes:

(1) different atmospheric models;

(2) an error in the gf values;

(3) the use of the LTE or non-LTE approximations;

(4) different base lists of lines;

(5) selection of atomic or ion lines;

(6) selection of lines observed in the disk center or for the Sun as a star;

(7) an accuracy of measurements of equivalent widths;

(8) a way of determining the micro- and macroturbulent velocities;

(9) a choice of the parameters of damping due to collisions;

(10) different computer codes.

Each of the listed causes of the appearing differences contributes
its effect to the result. For example, from the data of
\cite{2015A&A...573A..26S} (Table 3), it is seen that the
difference in the models yields the difference between the
abundance values from 0.04 to 0.11 dex. The large scatter of the
values of $A$ is produced by the uncertainty in the oscillator
strength. From the data of \cite{2011A&A...528A..87M}, the iron
abundance is 7.41--7.56 depending on the oscillator strength,
i.e., the maximum difference is 0.15 dex. The non-LTE effects
reduce the abundance by 0.01 dex (the data from
\cite{2012MNRAS.427...27B} in Table 3). As is shown by Holzreuter
and Solanki \cite{2013A&A...558A..20H}, the accounting for the
combined effect of the horizontal radiation transfer and the
non-LTE approximation on the iron lines in the realistic 3D
atmospheres will reduce the iron abundance by approximately 0.012
dex. The choice of the list of lines may cause a difference of
0.04 dex (the data from \cite{2006ApJ...636..821F,
2007ApJ...661.1152F} in Table 2).

It is also seen from Table 3 that the difference between the
values of the iron abundance derived from the analysis of the Fe I
and Fe II lines may be rather large, from 0.01 to 0.09 dex. It
depends on both the choice of the spectral lines and the
atmospheric model. In the cooler photospheres (e.g., MARCS), this
difference is 0.01--0.02 dex, while it reaches 0.05 dex in the
hotter ones (e.g., HM \cite{1974SoPh...39...19H}). In our case,
the difference between the values obtained from the Fe I and Fe II
lines is minimum (0.01 dex). The iron abundance obtained from the
observations of the disk center and for the Sun as a star may
differ to a little degree. For example, it follows from paper of
Caffau et al. \cite{2011SoPh..268..255C} that, for the Sun as a
star, the 3D models yield the abundance 0.02 dex lower than for
disk center.

 \begin{figure}[b]
 \centerline{\includegraphics    [scale=1]{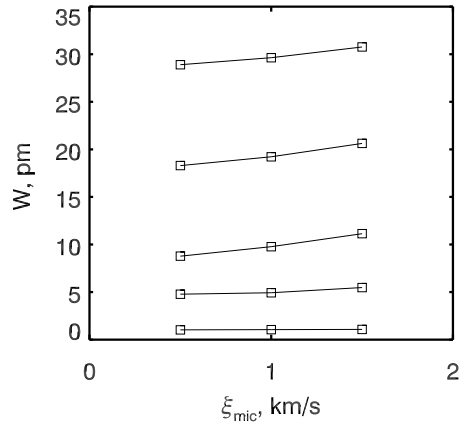}}
 \caption {\small
Change of the equivalent width $W$ in dependence on the parameter
of microturbulence $\xi_{\rm mic}$ for artificial absorption lines
of different strengths calculated for the Sun as a star.
  } \label{micro}
 \end{figure}

The abundance derived from the equivalent widths and 1D models
depends on the choice of the parameter of microturbulence. The
uncertainty in this parameter may lead to different equivalent
widths of the synthesized lines and, consequently, to different
values of the abundance.

Figure 4 demonstrates how the equivalent widths of the synthesized
lines (weak, moderate, and strong) may differ if the value of the
microturbulence parameter 1 km/s is changed within the limits of
$\pm0.5$~km/s. As is seen, the equivalent widths of very weak
lines are practically constant under such variations of
microturbulence. The situation becomes more problematic for
moderate lines (partially saturated lines). To choose the
microturbulence parameter properly is of key importance for
calculations of these lines in the analysis of the abundances. For
example, if a moderate line with the equivalent width $W \approx
5$~pm for the Sun as a star is synthesized with the value of 0.7
km/s instead of the standard value of 1 km/s, the derived
abundance will be 0.09 dex higher, and the difference will be even
larger for a moderate line with $W \approx 10$~pm.

The uncertainty in the microturbulence parameter, together with
the rotation velocity of the star, influences the shape of the
profiles and may introduce an error to the abundance value
obtained from the line profiles. According to our calculations for
weak lines, the increase of the macroturbulent velocity $\xi^{\rm
RT}_{\rm mac}$ from 3.0 to 3.5 km/s makes the abundance higher by
0.02 dex, while the twofold increase of the classic damping
constant due to collision processes yields a small increase of the
abundance, 0.01 dex. In the 3D models of the atmospheres, there is
no problem with micro- and macroturbulent velocities, though they
have their own problems and complexities. Moreover, it is not easy
to apply the 3D models to the analysis of the abundance in stellar
atmospheres, and their use for synthesizing the spectral lines is
time consuming.

If we sum (with no sign) the above-listed possible errors in the
estimate of the iron abundance, we will find the maximally
possible total error, 0.4--0.5 dex. This error is very large, and
it may appear only in rare cases when the base list of lines and
the atomic and atmospheric parameters are poorly selected. In real
practice, some errors are compensated, which results in a lower
total error.

Recently, Scott et al. \cite{2015A&A...573A..26S} have redefined
the chemical composition for the Sun with the use of the solar
spectrum in the disk center and the 3D models accounting for the
correction for deviation from LTE. For iron, they obtained the
value $A = 7.47 \pm 0.04$, and they recommend that it should be
used as the most reliable value currently available. They also
report the abundance value of $7.41 \pm 0.04$ obtained with the
MARCS models. In fact, both our and their values of the iron
abundance obtained with the 1D MARCS models of the atmosphere are
the same within the analysis errors. This confirms the reliability
of our result. Moreover, this allows us to estimate the correction
for horizontal inhomogeneity of the atmosphere that is ignored in
our analysis. Note, our method of analyzing the abundance excludes
systematic errors caused by inaccuracy in the equivalent widths,
non-LTE effects, and uncertainties in the micro- and
macroturbulent velocities and dumping constant. To  transform  the
abundance found by  Scott et al. \cite{2015A&A...573A..26S} ($A =
7.47$  for the solar center) to the abundance for the Sun as a
star, we introduce the correction of $-0.02$ dex and subtract our
abundance value ($7.42$). The result yields the highest possible
systematic error for atmospheric horizontal inhomogeneity, which
amounts to 0.03 dex. This means that the iron abundance determined
for the Sun as a star with the MARCS theoretical models of the
atmosphere is underestimated by 0.03 dex due to the 1D
approximation.

 \begin{table}
  \caption{\small
Macroturbulent velocity in the atmosphere of Arcturus and the Sun
as a star from the analysis of the Fe I and Fe II line profiles}
\vspace {0.5 cm}
  \label{T:comp3}
  \centering
  \footnotesize
  \begin{tabular}{l|l|l}
  \hline
\multicolumn{3}{c} { $ \xi^{\rm RT}_{\rm mac}$, km/s}
\\
\hline
  strong lines & weak lines  & Source\\
  \hline
 \multicolumn{3}{c} {Arcturus}\\
  $4.8\pm 0.2$ & $5.25\pm 0.2$ &    {Gray \cite{1981ApJ...245..992G}} \\
  $4.6\pm 0.3$ & --            &   {Sheminova \& Gadun \cite{1998KPCB...14..169S}} \\
  --           &$5.2\pm 0.2$   &   {Gray \& Brown \cite{2006PASP..118.1112G}} \\
  5.0          & --            &     {Tsuji \cite{2009A&A...504..543T}} \\
  --           &$5.6\pm 0.2$   &   This study \\
 \multicolumn{3}{c} {Sun as a star}\\
 $3.1\pm 0.1$  &  $3.8\pm 0.2$ &  {Gray \cite{1977ApJ...218..530G}} \\
 $2.6\pm 0.2$  &  --&  {Sheminova \& Gadun \cite{1998KPCB...14..169S}} \\
 $2.3\pm 0.4$  &4.0 & {Takeda \cite{1995PASJ...47..337T}} \\
  3.2  &--  &  {Gehren et al. \cite{2001A&A...366..981G}} \\
 --   &4.0   &  {Gehren et al. \cite{2001A&A...380..645G}} \\
  2.6  &  3.8&  {Mashonkina et al. \cite{2011A&A...528A..87M}} \\
 --& $3.45\pm 0.3$     & {Steffen et al. \cite{2013MSAIS..24...37S}}\\
 --&$3.5\pm 0.3$       &  This study \\
 \hline
 \end{tabular}
 \end{table}
 \noindent

As regards the macroturbulent velocity derived in the present
analysis for Arcturus, the values $ \xi^{\rm RT}_{\rm mac} = 5.57
\pm 0.24 $ and $5.60 \pm 0.2$~km/s (from the Fe I and Fe II lines,
respectively) do not contradict the data obtained earlier in the
Fourier analysis (Table 4). Gray \cite{1981ApJ...245..992G} found
the value $ \xi^{\rm RT}_{\rm mac} =5.25$~km/s. The fact that our
estimate is somewhat higher can be explained by two causes.

The first cause is that we used weaker lines. In the atmosphere of
Arcturus, the macroturbulent velocities decrease with height,
while the microturbulent velocities increase due to
defragmentation of large elements, lifting up. In the analysis of
strong lines, the retrieved velocities $\xi^{\rm RT}_{\rm mac}=
4.8$~km/s \cite{1981ApJ...245..992G} and $4.6 \pm 0.3$~km/s
\cite{1998KPCB...14..169S} are lower than ours.

The second cause is that we used a new value for the rotation
velocity of Arcturus  $V \sin i=1.5$~km/s
\cite{2006PASP..118.1112G}, while the other studies (see Table 4)
are based on the higher velocity $V \sin i=2.4$~km/s
\cite{1982ApJ...262..682G}. In the procedure of adjusting the
spectra, the decrease of the rotation velocity should influence
the parameter $ \xi^{\rm RT}_{\rm mac}$ by raising its value to
some extent. If the new value of the rotation velocity is correct,
we may recommend that the new value of macroturbulence $ \xi^{\rm
RT}_{\rm mac}=5.6$~km/s, together with the new value of rotation
$V \sin i=1.5$~km/s, should be used in the analysis of weak lines
of Fe I and Fe II.

Since the rotation velocity is a reliable parameter for the Sun,
the differences between the values of macroturbulent velocity
obtained by different researchers may be caused by the discrepancy
in the formation depth of the lines, from which the macroturbulent
velocity is determined. Table 4 contains the values of $ \xi^{\rm
RT}_{\rm mac}$ taken from different papers. As is seen, the values
of $ \xi^{\rm RT}_{\rm mac}$ obtained from strong lines are
smaller than those from weak lines. This is a well-known fact
indicating the decrease of $ \xi^{\rm RT}_{\rm mac}$ with height
in the photosphere. For all of the weak lines, we obtained the
mean value $\xi^{\rm RT}_{\rm mac}= 3.5 \pm 0.3$~km/s. Gray
\cite{1977ApJ...218..530G} found $\xi^{\rm RT}_{\rm mac}= 3.8 \pm
0.2$ and $3.1 \pm 0.1$~km/s for the lower and upper photosphere,
respectively. The decrease of $ \xi^{\rm RT}_{\rm mac}$ with
height is also confirmed by our data acquired from the Fe I ($3.22
\pm 0.09$~km/s) and Fe II ($3.69 \pm 0.11$~km/s) lines. From the
data for the center of the solar disk contained in the tables by
Gurtovenko and Sheminova \cite{1997MAO.....1P...3G}, we estimated
the mean height where these lines are formed. On the average, the
weak lines of Fe II are effectively formed at the level of 146 km,
while the weak lines of Fe~I are formed at the level of 182 km.
Because of this, the values of macroturbulent velocity in the
solar photosphere derived from the Fe~I and Fe~II lines differ.
Note that Arcturus does not show such difference in the values of
macroturbulent velocity obtained from the same weak lines of Fe I
and Fe II, since the formation conditions for these lines on
Arcturus differ from those on the Sun.

\section{Conclusions}

Arcturus is often used as a standard for studying the chemical
composition of the other similar giant stars. Fortunately, for
Arcturus, the spectrum of extremely high quality, from the UV
range to the IR one, is available together with the spectrum of
the Sun as a star (the atlas by Hinkle and Wallace
\cite{2005ASPC..336..321H}). Since Arcturus is a relatively close
star, the parameters of its atmosphere are known better than those
of any other RGB star. Because of this, it was of particular
interest to apply the method of sensitive points exactly to
Arcturus in order to estimate the iron abundance in its atmosphere
more accurately.

From the synthesis of weak spectral lines of Fe I and Fe II with
the MARCS models of the atmosphere, we determined the absolute
value of the iron abundance for Arcturus ($A = 6.95 \pm 0.03$) and
the Sun as a star ($A = 7.42 \pm 0.02$) in the usual logarithmic
scale relative to hydrogen. We also applied the differential
method to determining the iron abundance relative to the Sun and
derived [Fe/H]$~=-0.48 \pm 0.02$ on average.

It is especially important to know a high-accuracy value of the
absolute abundance of iron for the application of the differential
method to the analysis of the abundance in the other stars. As
Fulbright et al. \cite{2006ApJ...636..821F} note, not only the
compensation of errors in the oscillator strength is an advantage
of the differential analysis. In a first approximation, the errors
caused by different unaccounted effects in the atmospheric models,
such as 3D hydrodynamics, granulation, non-LTE, magnetic fields,
chromospheric effects, etc., may compensate each other when the
analysis is performed with this method relative to the similar
stars having rather close parameters of the atmosphere. Thus, for
many stars, the differential abundances can be obtained more
reliably than the absolute abundances based on the atmospheric
models.

For Arcturus, we recommend using the absolute value of the iron
abundance $A = 6.95 \pm 0.03$. We also recommend that the
sensitive points of the profiles of preferably weak lines should
be used instead of the full profile, especially in the cases when
the standard method of equivalent widths cannot be applied.

{\bf Acknowledgments.} We are grateful to C.R. Cowley for his
suggestion, realized in the present study, that the sensitive
points of the profiles are used for determining the abundance. We
would like to thank the reviewer for valuable comments.

\end{document}